\documentclass[a4paper,11pt]{article}
\usepackage{a4wide}

\usepackage{graphicx}
\usepackage{amsmath}
\usepackage{bm}

\newcommand{\pfrac}[2]{\frac{\partial{#1}}{\partial{#2}}}
\newcommand{\ppfrac}[2]{\frac{\partial^2{#1}}{\partial{#2}^2}}
\newcommand{\pppfrac}[2]{\frac{\partial^3{#1}}{\partial{#2}^3}}

\newcommand{\pilfrac}[2]{\partial_{#2} {#1}}

\newcommand{\pppilfrac}[2]{\partial^3_{#2} {#1}}

\newcommand{\ilfrac}[2]{{#1}/{#2}}
\newcommand{\lr}[1]{\left( {#1} \right)}
\newcommand{\lrsq}[1]{\left[ {#1} \right]}
\newcommand{\lrcur}[1]{\left\{ {#1} \right\}}
\newcommand{\lreval}[1]{\left. {#1} \right|}
\newcommand{\lrmod}[1]{\left| {#1} \right|}

\newcommand{\refe}[1]{(\ref{#1})}
\newcommand{\mye}{\mbox{e}}
\newcommand{\infinity}{\infty}

\newcommand{\dint}{\:\mathrm{d} }

\newcommand{\tbf}[1]{\textbf{#1}}

\newcommand{\nabS}{\bm{\nabla}_{\hspace{-0.1cm}s}}
\newcommand{\bmN}{\bm{\nabla}}
\newcommand{\bmDot}{\bm{\cdot}}

\newcommand{\etal}{\textit{et al.}}
\newcommand\ie{i.e.\ }
\newcommand\eg{e.g.\ }

\newcommand\vect[1]{\tbf{#1}}
\newcommand\tens[1]{\tbf{#1}}

\newcommand{\iT}{\tens{I}}

\newcommand{\tT}{\tens{T}}

\newcommand{\tbE}{\vect{e}}
\newcommand{\tbT}{\vect{t}}

\newcommand{\tbV}{\vect{v}}
\newcommand{\tbU}{\vect{u}}
\newcommand{\tbN}{\vect{n}}

\newcommand{\Bax}{{x}}
\newcommand{\Bat}{{t}}
\newcommand{\Bah}{{h}}

\begin{document}

\title{A comparison of slip, disjoining pressure, and interface formation models for contact line motion through asymptotic analysis of thin two-dimensional droplet spreading}

\author{David N. Sibley$^{1}$, Andreas Nold$^{1}$, Nikos Savva$^{2}$, and Serafim Kalliadasis$^{1}$
\\
\small{$^{1}$Department of Chemical Engineering, Imperial College London, London SW7 2AZ, UK}
\\
\small{$^{2}$School of Mathematics, Cardiff University, Cardiff CF24 4AG, UK}
}

\date{Published: J. Eng. Math. (2014); DOI: 10.1007/s10665-014-9702-9}

\maketitle

\begin{abstract}
The motion of a contact line is examined, and comparisons drawn, for a variety of models proposed in the literature. Pressure and stress behaviours at the contact line are examined
in the prototype system of quasistatic spreading of a thin two-dimensional droplet on a
planar substrate. The models analysed include three disjoining pressure models based on van der Waals interactions, a model introduced for polar fluids, and a liquid-gas diffuse-interface model; Navier-slip and two nonlinear slip
models are investigated, with three microscopic contact angle boundary conditions imposed (two of these contact angle conditions having a contact line velocity dependence); and the interface formation model is also considered.
In certain parameter regimes it is shown that all of the models predict the same quasistatic droplet spreading behaviour.

\vspace{1em}

\noindent\textit{Keywords:} Contact line, slip, diffuse-interface, precursor film, disjoining pressure, interface formation.
\end{abstract}

\section{Introduction}
Immiscible fluids and non-porous solids occur abundantly in both nature and industry. The location at which two such fluids and a solid meet is termed a contact line, and the modelling of moving contact lines is consequently of central importance to the understanding of problems as diverse as how insects walk on water and the wetting properties of plant leaves to coating, inkjet printing, and oil recovery---for reviews, see \eg \cite{Dussan79,deGennesrev,blake2006physics,BonnEggers,SnoeijerRev,SuiDingSpeltRev}. The main focus of this work is to compare a number of prevalent models proposed to resolve the \emph{moving contact line problem} in the prototype system of quasistatic spreading of a thin two-dimensional droplet on a planar substrate.

Two early papers sparked interest in the moving contact line problem. The first by Moffatt in 1964 \cite{Moffatt} considered the conceptually simple problem of a wedge of viscous fluid in contact with an inviscid fluid (the interface kept flat ostensibly by the action of gravity), and also in contact with a solid substate moving with a constant velocity $U_w$. In this situation, Moffatt noted that in the absence of fluid inertia, a fluid particle initially on the fluid interface will be moving towards the contact line at a speed independent of the distance away from it, and with a value less that $U_w$. The no-slip boundary condition will then enforce the speed of this fluid particle to instantaneously increase to $U_w$ at the contact line, where it `turns the corner'. This would require an infinite acceleration caused by an infinite stress and pressure at the contact line, all of which are physically unacceptable.

This relatively brief glimpse into the moving contact line problem was brought into focus by the seminal work in 1971 of Huh and Scriven \cite{HuhScriv71}. The authors extended the analysis of Moffatt to include the situation of two immiscible viscous fluids of arbitrary viscosities and described in detail the \emph{nonintegrable shear-stress singularity} alongside offering suggestions as to how it may be overcome. Huh and Scriven highlighted the primary cause for the unphysical singularities as the no-slip condition at the wall. To resolve the moving contact line problem,
Huh and Scriven suggested that some form of slip in the contact line vicinity could be allowed, such
as the Navier-slip condition initially proposed in the far earlier work by Navier \cite{Navier}. Whilst this slip boundary condition was not initially proposed with a physical basis, it should
be noted that in the last few decades there is evidence that slip can physically occur at the molecular scale, such as seen in the molecular dynamics simulations of \cite{thompsontroian97,HeHadjiJFM,QianWangShengGNBCfirst}, and there are numerous physical effects such as viscous heating, dissolved gas or polymer solutions, and surface roughness that may contribute to apparent slip \cite{LaguaBrennerStone}.

Since these important works, including the work of Dussan V. and Davis~\cite{DussanDavis} who in 1974 clarified the moving contact line problem mathematically and with experiments, the volume of literature on the moving contact line problem has increased rapidly, as evidenced in the subsequent review articles~\cite{Dussan79,deGennesrev,blake2006physics,BonnEggers,SnoeijerRev,SuiDingSpeltRev}. The problem remains an active field of research, due in part to the wide variety of microscale physical effects which can resolve the problem. Further to the above, these include modelling a precursor film ahead of the contact line (physically realised through the inclusion of a disjoining pressure, which accounts for intermolecular interactions near the contact line)~\cite{SchwartzEley}; rheological (\ie non-Newtonian) effects~\cite{WeidnerSchwartz}; treating
surfaces as separate thermodynamic entities with dynamic surface tensions (as in the interface formation model)~\cite{Shikh93}; introducing numerical slip~\cite{RenardySlip}; assuming the contact angle to be always $180^\circ$ at the solid substrate~\cite{benilov180}, including evaporative fluxes~\cite{redcolevap}; considering the interface to be diffuse~\cite{anderson_rev}---with different models for liquid-gas~\cite{Seppecher} and liquid-liquid (binary) fluids~\cite{jacqmin} being adopted; and modelling both molecular and hydrodynamic effects in hybrid models~\cite{Petrov2Hybrid}.

In the present communication we will compare a variety of models that are used to alleviate the non-integrable
stress singularity at a moving contact line, concentrating primarily on slip and precursor film models (including one
derived from diffuse-interface theory) as well as the aforementioned interface formation model. In slip models, the no-slip boundary condition is relaxed to allow for fluid particles to move along the substrate. On the other hand, the no-slip condition is applied on the substrate for
precursor film models, but has only an apparent contact line away from the substrate, thus removing the associated
stress and pressure singularities that would have occurred had the contact line been sharp. As the substrate must be entirely covered by a film in these models a disjoining pressure is included due to long-range intermolecular interactions
(such as from van der Waals forces \cite{Waals_trans,Waals_orig}), with both repulsive and attractive contributions to this pressure balancing to create a stable thin-film. Details of the interface formation model will be given in Sec.~\ref{sec:thinfilm} where it is analysed.

Alongside the question of which microscale effects to include in a description of the moving contact line, for many models there is another related question as to which additional boundary condition(s) at the contact line should be imposed.
For slip models, the boundary condition at the contact line usually takes the form of a prescription of the variation of contact angle with contact line velocity. A simple, yet widely used condition is
that the microscopic contact angle remains at its static value when the
contact line is in motion, \ie $\theta_c = \theta_d = \theta_S$, where $\theta_c$ is the microscopic contact angle in any situation, $\theta_d$ is the dynamic contact angle, and $\theta_S$ is the static contact angle. This is imposed in a number of studies involving slip models such as studies by Hocking \cite{Hocking83} and Cox
\cite{CoxPart1}, as well as a recent analysis providing model comparisons for 2D droplet spreading by Savva and Kalliadasis \cite{SavvaPrecursorSlip}. The analysis of other contact angle conditions and models to extend the work of \cite{SavvaPrecursorSlip}, given in Sec.~\ref{sec:thinfilm} here, is the primary focus of this article.
Two forms we consider here to allow for contact line velocity dependence on microscopic contact angle are
\begin{align}
 U_w = \bar{k}(\theta_d-\theta_S)^m, \qquad \mbox{or} \qquad U_w = \bar{k}(\theta_d^m-\theta_S^m), \label{eq:introCArel}
\end{align}
where $\bar{k}$ is a coefficient with dimensions of velocity, and $m$ is a dimensionless exponent. The parameters $\bar{k}$ and $m$ could be determined from experiments (see, for instance, \cite{Hoffman,FermigierJenffer,RameGaroff,Sikalo} for some experimental measurements) but the question as to whether the contact angle varies dynamically \emph{at} the substrate remains an open debate, given the limits of experimental resolution---see discussion relating to this in \cite{My_Shikh}. Forms such as \refe{eq:introCArel} are considered, with $m$ specified, for instance by Greenspan \cite{greenspan}, Haley and Miksis \cite{HaleyMiksis}, Hocking \cite{HockingRival}, Ehrhard and Davis \cite{ehrharddavis}, and Wilson \etal~\cite{wilson_spin}.

For the disjoining pressure models considered here, where there is an apparent contact line due to a precursor film, a different contact angle relation may be derived. For long-wave models (where $\theta_S\ll1$) the contact angle is given by
\begin{align}
 \theta_S^2 = -2 V(H^*)/\sigma_{12} , \qquad \bar\Pi(H) = \pfrac{V}{H}, \label{eq:earlyangledp}
\end{align}
where $H$ is the height of the free surface, $H^*$ is the precursor film thickness,
$\sigma_{12}$ is the fluid-fluid surface tension, $V(H)$ is the interface potential for a flat film of thickness $H$, and $\bar\Pi(H)$ is the disjoining pressure, which here depends on the thin-film height. This long-wave equilibrium contact angle relation \refe{eq:earlyangledp} is derived through multiple integrations of the equilibrium thin-film equation for precursor film/disjoining pressure models (\eg by Eggers \cite{EggersPRE05}), and is also used in the comparison study \cite{SavvaPrecursorSlip}. We note that \refe{eq:earlyangledp} is not an imposed boundary condition, since the effect of the disjoining pressure creates a precursor film and prevents an actual contact line occurring. The required boundary conditions are instead given by $H$ approaching the precursor film thickness $H^*$ at distances far from the bulk liquid region.

Before discussing these models and conditions in the thin-film (long-wave) approximation in Sec. \ref{sec:thinfilm}, we revisit the early work of Moffatt \cite{Moffatt}, which imposed no-slip, in Sec. \ref{sec:mofhuhscriv} to uncover the nature of the moving contact line singularity. We then extend the analysis to see how slip models overcome the problem in this simple situation, and to show that no essential physics is missing in the thin-film setting. The results will be in agreement with the later thin-film analysis of Sec. \ref{sec:thinfilm}.

\section{Moffatt wedge flow}
\label{sec:mofhuhscriv}

Whilst the problems considered by Moffatt \cite{Moffatt} and Huh and Scriven \cite{HuhScriv71} are unphysical---as in the imposed geometry the normal stress balance condition at the fluid-fluid free surface is not satisfied, it is instructive to revisit the situation and consider how slip conditions change the analysis.

\begin{figure}[ht]
\centering
\includegraphics[width=11cm]{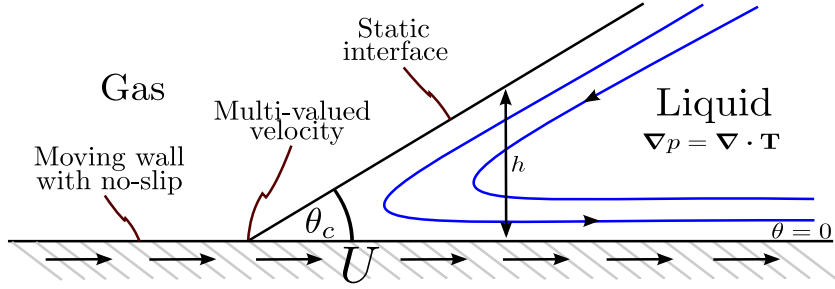}
\caption{Model of contact line motion in a wedge geometry with contact angle $\theta_c$, working in a frame of reference where the contact line is stationary and the solid is moving with nondimensional velocity $U$.}
\label{fig:fig1_HS_wedge}
\end{figure}

Consider the motion of a contact line on a perfectly flat solid surface, with
governing equations appropriately nondimensionalised. It is convenient to
work in a frame of reference where the contact line is fixed and the surface
translates with steady (nondimensional) velocity $U$. The contact line is the
location where the fluid free surface meets the solid, the free surface is
assumed to be planar at an angle $\theta=\theta_c$.
The situation is two-dimensional and we take the contact line to be the
origin of a polar coordinate system. The situation is shown in Figure
\ref{fig:fig1_HS_wedge}.

For rapid wetting processes inertial effects may be important, and have been studied analytically \cite{CoxHighRe}, and recently validated numerically \cite{SuiJFM13}. Throughout this article however, we will consider inertialess flow relevant for slow processes such as droplet spreading. In the Moffatt wedge flow geometry, this means we consider the creeping (or Stokes, \cite{Stokes}) flow regime, where the nondimensional governing equations for inertialess incompressible Newtonian flow are
\begin{align}
\bmN\bmDot\tbU = 0, \qquad \bmN p = \bmN\bmDot\tT, \qquad
\tT = \bmN\tbU + (\bmN\tbU)^\mathrm{T} , \qquad
\tbU = u\tbE_r + v\tbE_\theta = \frac{1}{r}\pfrac{\psi}{\theta}\tbE_r -\pfrac{\psi}{r}\tbE_\theta,
\label{eq:GEnewtonian}
\end{align}
where $\tbU$ is the fluid velocity, $\tT$ is the extra stress tensor, $p$ is the pressure, and $\psi(r,\theta)$ is the stream function in polar coordinates.
Consider the behaviour as $r\to0$ of $\psi=O(r^{\gamma})$, with $\gamma$ an exponent to be determined. From this behaviour we can determine the order-of-magnitude estimates
\begin{align}
\psi = O(r^{\gamma}), \quad \tbU = O(r^{-1+\gamma}), \quad \tT = O(r^{-2+\gamma}),\quad \bmN\bmDot\tT = \bmN p = O(r^{-3+\gamma}), \label{eq:rscal}
\end{align}
so that the pressure order-of-magnitude is
\begin{align}
 p = O(r^{-2+\gamma}), \quad \mbox{for } \gamma \neq 2, \qquad \mbox{and} \qquad p = O(\log r), \quad \mbox{for } \gamma = 2 .
 \label{eq:presordwlog}
\end{align}

The momentum equation is easily reduced
to the biharmonic equation
$\nabla^4 \psi = 0$,
after introducing the stream function,
and we note the form of the pressure and stresses in polar coordinates, using \refe{eq:GEnewtonian}, may be found from
\begin{gather}
 \pfrac{p}{r} = \frac{1}{r^2}\frac{\partial^2 \psi}{\partial r \partial \theta} + \frac{1}{r}\frac{\partial^3 \psi}{\partial r^2 \partial \theta} + \frac{1}{r^3}\frac{\partial^3 \psi}{\partial \theta^3},
 \qquad
 \pfrac{p}{\theta} = -r\frac{\partial^3\psi}{\partial r^3} - \frac{\partial^2 \psi}{\partial r^2} - \frac{1}{r}\frac{\partial^3 \psi}{\partial r \partial \theta^2}+\frac{1}{r}\pfrac{\psi}{r}+\frac{2}{r^2}\ppfrac{\psi}{\theta},
\nonumber\\
T_{rr} = 2\pfrac{u}{r}, \qquad
T_{r\theta} = \pfrac{v}{r}+\frac{1}{r}\pfrac{u}{\theta}-\frac{v}{r}, \qquad
T_{\theta\theta} = \frac{2}{r}\left({u}+\pfrac{v}{\theta}\right).\label{eq:stressinpolar}
\end{gather}

Following Moffatt \cite{Moffatt}, we take a separable stream function
\begin{align}
\label{eq:newton:psib}
\psi = r^{\gamma} f_0 (\theta).
\end{align}
We note here that the power-law form \refe{eq:newton:psib} for the stream function is used, as in \cite{KirkinisDavis} who looked at other more complex forms of slip. This is in contrast to the analysis of Hocking \cite{Hocking77} for Navier-slip, who considered a form consisting of the no-slip solution modified by a logarithm. Hocking's analysis gives a full matched asymptotic solution, which yields complex integral expressions that do not afford easy insight into the nature of the stress and pressure behaviours. We instead focus our simple analysis on the behaviour at the contact line, essentially the limit as the contact line is approached from the Hocking inner solution---which becomes $O(r^2)$ for the Navier-slip model, as seen from equations (2.10), (2.11) and (2.15) there.

Returning to \refe{eq:newton:psib}, on substituting this into the biharmonic equation we obtain the linear ordinary differential equation (ODE) for $f_0$ of
\begin{align}
 f_0'''' + 2(\gamma^2-2\gamma+2)f_0'' + \gamma^2(\gamma-2)^2f_0 =0,
\end{align}
true for all $\gamma$, including (in particular) $\gamma=0$, $\gamma=1$, and $\gamma=2$. This is a fourth order ODE for $f_0$, with solutions
\begin{align}
  f_0 &= \left\{ \begin{array}{ll}
  C_1\sin[(\gamma-2)\theta]+C_2\cos[(\gamma-2)\theta]+C_3\sin(\gamma\theta)+C_4\cos(\gamma\theta),
  & \mbox{for }\gamma \neq \{0,1,2\},\\
  C_1+C_2\theta+C_3\sin(2\theta)+C_4\cos(2\theta), & \mbox{for }\gamma = \{0,2\},\\
  C_1\sin\theta+C_2\cos\theta+C_3\theta\sin\theta+C_4\theta\cos\theta,  &\mbox{for }\gamma = 1,
\end{array}\right.\label{eq:bhedeg}
\end{align}
noting the correction for the degenerate solution for $\gamma=0$ (in comparison to the erroneous form in \cite{Moffatt}).
The boundary conditions we apply in this (artificial planar wedge) configuration are of vanishing normal component of velocity at the solid surface and the fluid interface
\begin{align}
 v = 0 \quad \mbox{on} \quad \theta=0, \qquad v=0\quad  \mbox{on}\quad  \theta = \theta_c, \label{eq:bcnonv}
\end{align}
and the continuity of tangential stress at the interface
\begin{align}
\tbN\cdot\tT\cdot\textbf{t} = -\nabS\sigma\cdot\textbf{t}, \qquad \mbox{on}\quad \theta=\theta_c,\label{eq:bctanst}
\end{align}
where $\sigma$ is the (dimensionless) surface tension, $\tbN=\tbE_\theta$ is the unit normal pointing out of the fluid, and $\tbT=\tbE_r$, is a tangent vector on the interface.
The surface tension in the classical model is usually assumed to be a constant unless temperature variation or variable chemical composition are to be modelled. In this section we will also assume constant surface tension, so that the boundary condition becomes
\begin{align}
 T_{r\theta} = 0,\qquad \mbox{on}\quad \theta=\theta_c, \label{eq:bcts}
\end{align}
but will return to discuss this assumption in Appendix~\ref{app:wedge}.

The final condition is applied on the wall,
where the various slip models may be introduced. In classical fluid mechanics problems, no-slip is imposed on an interface between a viscous fluid and a solid boundary. For a moving contact line, this leads to a singular stress and pressure at the contact line, as discussed. Navier-slip, discussed in the introduction, has been used extensively,
and is of the form $(\tbU-\tbU_w)\bmDot\tbT = \lambda \tbN\cdot\tT\cdot\textbf{t}$. In this representation the variables are nondimensional, with $\tbU_w$ the wall velocity and $\lambda>0$ the nondimensional slip-length.
The possibilities for slip to overcome this problem we will investigate are:
\begin{align}
\left.
\begin{array}{ll}
 \mbox{No-slip} &u=U, \\ \mbox{Navier-slip} & u = U+\lambda T_{r\theta}, \\ \mbox{Inverse linear slip} & u = U+(\ilfrac{\lambda^2}{h}) T_{r\theta}, \\ \mbox{Inverse quadratic slip} \quad& u = U+(\ilfrac{\lambda^3}{h^2}) T_{r\theta},
\end{array}\right\}\quad \mbox{on} \quad \theta=0.\label{eq:slipmods}
\end{align}
The final two models have the slip length related to the height of the fluid region, with $h=r\sin\theta_c$, see Figure \ref{fig:fig1_HS_wedge}, and for these two models clearly we are restricted to contact angles $\theta_c<90^\circ$.
Whilst less used, the final two models do appear in the literature. In relation to the droplet spreading problem which this work will focus on, noteworthy are works by Greenspan \cite{greenspan} who used a form of the `inverse linear slip' condition, and the work of Haley and Miksis \cite{HaleyMiksis}, who employ all three above slip conditions.
In the long-wave limit, the `inverse linear slip' boundary condition reduces to a
slip condition proposed by Ruckenstein and Dunn \cite{RuckDunn}, which scales with the pressure gradient, as shown in \cite{SavvaPrecursorSlip}.

\subsection{Asymptotic behaviour and solutions}
We now consider the asymptotic behaviour of the various slip models given in \refe{eq:slipmods}. This should then allow us to choose an appropriate form for the stream function, and thus the general solution of the biharmonic equation for each model. We use the form of $T_{r\theta}$ from \refe{eq:stressinpolar}, with the stream function from \refe{eq:GEnewtonian}, and the separable form of the stream function in \refe{eq:newton:psib} to obtain
\begin{align}
 \left.
\begin{array}{ll}
 \mbox{No-slip} & r^{\gamma-1}f'_0=U, \\ \mbox{Navier-slip} &  r^{\gamma-1}f'_0 = U+\lambda r^{\gamma-2}\left(\gamma(2-\gamma)f_0+f_0''\right), \\ \mbox{Inverse linear slip} & r^{\gamma-1}f'_0 = U+(\ilfrac{\lambda^2 r^{\gamma-3}}{\sin\theta_c}) \left(\gamma(2-\gamma)f_0+f_0''\right), \\ \mbox{Inverse quadratic slip} \quad& r^{\gamma-1}f'_0 = U+(\ilfrac{\lambda^3 r^{\gamma-4}}{\sin^2\theta_c})\left(\gamma(2-\gamma)f_0+f_0''\right).
\end{array}\right\}
\quad \mbox{on} \quad \theta=0.
\label{eq:slipmodsr}
\end{align}
Consider \refe{eq:slipmodsr} now to determine $\gamma$. In the no-slip case, the only appropriate balance is when $\gamma=1$, giving a stream function of $\psi = r f_0 (\theta)$. In all other cases (\ie the slip cases), there is an apparent choice of balances. However, there is only one dominant balance. The left hand side term of \refe{eq:slipmodsr} which comes from $u=r^{\gamma-1}f'_0$ clearly cannot balance with the $T_{r\theta}$ terms (the terms involving $\lambda$ on the right hand side of \refe{eq:slipmodsr}), and attempting the possibility of balancing the $u$ term with the $U$ term at leading order would always leave the $T_{r\theta}$ terms dominating. Thus the dominant balance is between the $U$ and $T_{r\theta}$ terms, giving
\begin{align}
\left.
\begin{array}{lll}
 \mbox{No-slip} & {\gamma=1}, & \psi = r \left[  C_1\sin(\theta)+C_2\cos(\theta)+C_3\theta\sin(\theta)+C_4\theta\cos(\theta) \right], \\
 \mbox{Navier-slip} & {\gamma=2}, & \psi = r^2 \left[ C_1+C_2\theta+C_3\sin(2\theta)+C_4\cos(2\theta) \right], \\
 \mbox{General slip} & {\gamma=2+n}, &
 \psi = r^{2+n}\left[C_1\sin(n\theta)+C_2\cos(n\theta) \right. \\
 \quad(n>0) & & \left.\qquad\qquad\qquad\qquad\quad+\; C_3\sin((2+n)\theta)+C_4\cos((2+n)\theta) \right],
\end{array}\right\}
\label{eq:psiforms}
\end{align}
where the appropriate form for $f_0$ has been used from \refe{eq:bhedeg}, and the inverse linear and inverse quadratic slip models have been combined into a `general slip' model, with $n=1$ and $n=2$ corresponding to the inverse linear and inverse quadratic slip models respectively.

Using the appropriate stream function form from \refe{eq:psiforms}, we may now use the boundary conditions \refe{eq:bcnonv}, \refe{eq:bcts} and the desired slip model from \refe{eq:slipmods} to determine the stream function solution, and thus the velocity, stress and pressure forms. With the no-slip condition applied on the wall, the solution satisfying the boundary conditions is
\begin{gather}
\psi \sim {rU}{Q_c^{-1}}\left( \theta_c\sin\theta - \theta\sin\theta_c\cos(\theta_c-\theta) \right),
\nonumber\\
u \sim {U}Q_c^{-1}\left(
\theta_c\cos\theta-\sin\theta_c(\theta\sin(\theta_c-\theta) + \cos(\theta_c-\theta))
\right),
 \qquad
 v \sim -r^{-1}{\psi},
\qquad
 T_{rr} = T_{\theta\theta} = 0, \nonumber\\
 T_{r\theta} \sim -{U}(r Q_c)^{-1}\left( 2\sin\theta_c\sin(\theta_c-\theta) \right),
\qquad
p \sim - {U}(r Q_c)^{-1}\left( 2\sin\theta_c\cos(\theta_c-\theta) \right),\label{eq:no-slip_sol}
\end{gather}
where $Q_c = \theta_c-\sin\theta_c\cos\theta_c$.
If we now consider the velocity on the free surface $\theta=\theta_c$, it is easy to see that $v=0$, as required, and that
\begin{align}
 u &\sim \frac{U\left(
\theta_c\cos\theta_c-\sin\theta_c
\right)}{\theta_c-\sin\theta_c\cos\theta_c}
   = -U\lrsq{1 - \frac{
(\theta_c-\sin\theta_c)(1 + \cos\theta_c)
}{\theta_c-\sin\theta_c\cos\theta_c}},
\end{align}
giving the solution of Moffatt \cite{Moffatt}---see equation (2.3) there, and confirming that the free surface velocity is a monotonically decreasing function of $\theta_c\in[0,\pi]$, having values in the range $u\sim-U/2$ to $u\sim-U$.

The $r^{-1}$ singularities are clearly seen in \refe{eq:no-slip_sol} in the shear stress and pressure, but this result is also of interest to motivate why taking the contact angle as $180^\circ$ (not excluded for this no-slip case, and so that the liquid advances and rolls as would a caterpillar track) could remove the singularity and be worthy of investigation---at this order the above equations for $\theta_c=180^\circ$ give $O(1)$ velocities but zero pressure and stresses.
Analysis of this special case was considered by Benney and Timson \cite{benneytimson} in 1980, but a sign error was later found \cite{ngandussan}. The analysis has since been revisited by Benilov and Vynnycky in 2013, \cite{benilov180}, for the problem of two-dimensional Couette flow with a free boundary. However here we focus on small contact angles, which allow us to look into models applied for thin film equations.

The solutions for the different slip models in the Moffatt wedge setting may be found in Appendix~\ref{app:wedge}, however the important results are that
the Navier-slip model removes the singularity in the stress as the contact line is approached, but a logarithmic singularity in the pressure remains. This is integrable, however, and as such is not as worrying as the $1/r$ singularity in the pressure and stress in the no-slip model. In comparison, the two cases $n=1$ and $n=2$ in the general slip model resolve both the pressure and stress singularities, but both give zero pressure and stresses as the contact line is approached---the pressure and stresses being $O(r)$ and $O(r^2)$ in the two cases respectively. Finally shown in Appendix~\ref{app:wedge} is a solution with $O(1)$ stresses, yet without the logarithmic singularity in the pressure, which is achieved through allowing for variable surface tension.

\section{Thin-film regime}
\label{sec:thinfilm}

A useful and illuminating set of flows to consider the moving contact line in
are that of thin-films \cite{BertozziNAMS,oron1997long}.
In the thin-film approximation
the characteristic scale in the direction normal to the substrate is much smaller than that of the direction tangential to it. The small parameter of the system, which may be either purely the ratio of the normal to tangential scales, or based on (for instance) the smallness of the contact angle, then allows many of the terms in the full Navier--Stokes equations and boundary conditions to be neglected, with analytical progress becoming more feasible.

Here, we will consider contact line motion in the prototype setting of the
spreading of a thin two-dimensional droplet, as in previous works
\cite{Hocking83,Savva09,My_Shikh}, noting that the analysis holds also
for receding droplets without modification, as seen in
\cite{Savva09,My_Shikh}. For motion outside the validity of our quasistatic
assumption, however, spreading and receding droplets may have to be treated
differently due to the different asymptotics of the full inner-region
equation \cite{DuffyWilson,eggershigherCa}. Ideally, the full equations are
considered in dimensional form and the small terms are neglected after a
process of nondimensionalisation (as performed in the thin-film reviews
\cite{oron1997long,craster_matar} and also \cite{kalliadasis_thiele} for slip
and precursor film models, to which the reader is directed for full details),
noting also a relatively new, yet elegant and powerful method based on
gradient dynamics where a variational procedure allows the time evolution of
the film height to be derived as a conserved order parameter \cite{epj10}.
Slip is accounted for in the boundary condition at the wall, whereas
intermolecular forces due to the wall are included through a disjoining
pressure in the Navier--Stokes equations~\cite{WilliamsDavis}. Unlike the
critique formulated in~\cite{epj1}, the disjoining pressure does not
`double-count' intermolecular forces of the fluid as it only accounts for the
additional effects due to the presence of the wall-fluid interface. We
do note, however, that there remains some debate about the issue of including
additional effects from the microscale in continuum mechanics equations, as
also discussed in~\cite{ShikhBook}.

For completeness here we will derive the evolution equation from the two-dimensional Navier--Stokes equations already reduced to the leading order long-wave form \cite{oron1997long,craster_matar,ockock}. As mentioned, it is assumed that the characteristic scale in the direction normal to the substrate is much smaller than that of the direction tangential to it. Lubrication theory is known to be a good approximation for slowly moving, thin, viscous droplets \cite{greenspan,Hocking83,oron1997long}. The equations reduce to the continuity equation and simplified momentum equations where the pressure gradient (with disjoining pressure gradient optionally included) is non-zero in the $\bar{x}$-direction only and balanced by the vertical shear of the horizontal velocity.
Thus the (dimensional) governing equations are
\begin{align}
 \pfrac{\bar{u}}{\bar{x}} + \pfrac{\bar{v}}{\bar{z}} = 0, \qquad
 \pfrac{\bar{p}}{\bar{z}} = 0, \qquad
 \pfrac{\bar{p}}{\bar{x}} + \frac{\dint \bar{\Pi}}{\dint H}\pfrac{H}{\bar{x}} =  \mu \ppfrac{\bar{u}}{\bar{z}},
 \label{eq:lwgov}
\end{align}
with boundary conditions
\begin{subequations}
\begin{align}
 \lreval{\bar{u}}_{\bar{z}=0} = \frac{\bar{\lambda}^{1+n}}{H^n}\lreval{\pfrac{\bar{u}}{\bar{z}}}_{\bar{z}=0} , \qquad
 \lreval{\bar{v}}_{\bar{z}=0} = 0, \qquad
 \lreval{\pfrac{\bar{u}}{\bar{z}}}_{\bar{z}=H} = 0, \label{eq:lwbc1}
 \\
 \lreval{-\bar{p}}_{\bar{z}=H} + \bar{p}_a = \sigma\lreval{\ppfrac{H}{\bar{x}}}_{\bar{z}=H} , \qquad
 \pfrac{H}{\bar{t}} + \lreval{\bar{u}}_{\bar{z}=H}\pfrac{H}{\bar{x}} = \lreval{\bar{v}}_{\bar{z}=H},\label{eq:lwbc2}
\end{align}
\end{subequations}
which are the thin-film forms of the tangential stress balance, slip law, no penetration condition, normal stress balance, and kinematic (free-surface) boundary condition, respectively.\footnote{We note that in equations \refe{eq:lwgov}--\refe{eq:lwbc2} we have included the disjoining pressure in the momentum equation, as in \cite{WilliamsDavis,oron1997long,MunchWagnerWitelski}. A derivation may be made with the disjoining pressure instead entering into the normal stress balance in eq.~\refe{eq:lwbc2}(i), as \eg \cite{ajaev,FuentesCerroNewModel}, however the same evolution equation in \refe{eq:dimlw} is then found.}
Now, by using \refe{eq:lwgov}(ii) and \refe{eq:lwgov}(iii), the velocity $\bar{u}$ may be determined in terms of $\bar{p}=\bar{p}(\bar{t},\bar{x})$, $\bar{\Pi}$ and $H$, with $\bar{v}$ then determined through \refe{eq:lwgov}(i). The three arbitrary functions involved in $\bar{u}$ and $\bar{v}$ are then determined with the three boundary conditions \refe{eq:lwbc1}, with the pressure then given using \refe{eq:lwbc2}(i). Finally, all of the results are substituted into the kinematic boundary condition \refe{eq:lwbc2}(ii) yielding the evolution equation of the film height
\begin{align}
 \pfrac{H}{\bar{t}} + \frac{1}{3\mu}\pfrac{}{\bar{x}}\lrcur{ H^2 \lr{H + 3\frac{\bar{\lambda}^{1+n}}{H^n}}
 \pfrac{}{\bar{x}}\lrsq{ \sigma\ppfrac{H}{\bar{x}} - \bar{\Pi}(H) } }=0. \label{eq:dimlw}
\end{align}
We note that \refe{eq:dimlw} with $n=0$ is essentially the thin-film equation in the `weak-slip regime' of \cite{MunchWagnerWitelski}, which also considered a number of other nondimensionalisations for circumstances where slip plays a greater role. A number of other similar thin-film equations, predominately when the substrate is covered by a precursor film, have been considered using asymptotics in \cite{KingBowen}.

There will be three important quantities to consider at the contact line---the pressure, the velocity parallel to
the substrate, and the shear stress. In the long-wave regime (and at $\bar{z} = 0$), these correspond to
\begin{gather}
 \mbox{Pressure:} \quad \bar{p} = \bar{p}_a - \sigma\ppfrac{H}{\bar{x}}, \qquad
 \mbox{Velocity:} \quad \lreval{\bar{u}}_{\bar{z}=0} = \frac{\sigma\bar\lambda^{1+n}{H}^{1-n}}{\mu}\pppfrac{H}{\bar{x}}, \nonumber\\
 \mbox{Shear stress:} \quad \lreval{\pfrac{\bar{u}}{\bar{z}}}_{\bar{z}=0} = \frac{\sigma{H}}{\mu}\pppfrac{H}{\bar{x}}.
\label{eq:dim_pvs}
\end{gather}

The contact angle $\theta_c$ in general is given by $\tan\theta_c = \pilfrac{H}{\bar{x}}$, but as it must be small it may be used as the long-wave parameter, with the result $\theta_c\sim\pilfrac{H}{\bar{x}}$. We nondimensionalise eqs.~\refe{eq:dimlw}--\refe{eq:dim_pvs} through the scalings
\begin{gather}
 \bar{x} = L x, \qquad \bar{t} = \frac{3\mu L}{\sigma} t, \qquad H = L h, \qquad
 \bar{u} = \frac{\sigma}{3\mu} u, \qquad
 \bar{p} = \bar{p}_a + \frac{\sigma}{3L} p, \nonumber\\
 \bar{\lambda} = \frac{L}{3^{1/(1+n)}}\lambda, \qquad \bar{\Pi}(H) = \frac{\sigma}{L}\Pi(h),\label{eq:ndscalsdp}
\end{gather}
which gives the evolution equation
\begin{align}
 \pfrac{h}{t} + \pfrac{}{x}\lrcur{ h^2 \lr{h + \frac{\lambda^{1+n}}{h^n}}
 \pfrac{}{x}\lrsq{ \ppfrac{h}{x} - \Pi(h) } }=0, \label{eq:ndlw}
\end{align}
and
\begin{align}
 p = -3\ppfrac{h}{x}, \qquad
 u|_{z=0} = \lreval{\lrsq{\lambda^{1+n}h^{1-n}\pppfrac{h}{x}}}_{z=0}, \qquad
 \lreval{\pfrac{u}{z}}_{z=0} = \lreval{\lrsq{3 h \pppfrac{h}{x}}}_{z=0}.\label{eq:nd_pvs}
\end{align}

\begin{figure}[ht]
\centering
\includegraphics[width=12cm]{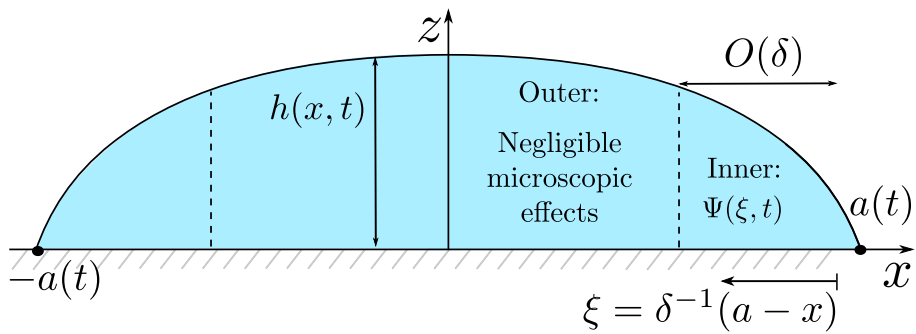}
\caption{The droplet geometry with asymptotic regions shown: for all models considered the microscopic effects such as slip or disjoining pressure are negligible in the outer region, but significant in $O(\delta)$ inner regions near the contact lines, located at $x=\pm a(t)$.}
\label{fig:drop}
\end{figure}

For slip models here, we will restrict our attention to the three cases discussed by Haley and Miksis \cite{HaleyMiksis} of $n=\{0,1,2\}$. In \cite{SavvaPrecursorSlip}, the authors described how the Navier-slip and inverse linear slip conditions gave equivalent spreading dynamics as precursor film models, with two forms being investigated based on the models used by Schwartz and Eley \cite{SchwartzEley}, and by Sharma \cite{Sharma}. These two disjoining pressure models correspond to
\begin{align}
 \bar\Pi(H) &= \frac{A^*}{H^{*m^*}}\lrsq{\lr{\frac{H^*}{H}}^{m^*} - \lr{\frac{H^*}{H}}^{n^*}}, \nonumber\\
 \bar\Pi(H) &= \frac{A^*}{H^{*m^*}}\lrsq{\lr{\frac{H^*}{H}}^{m^*} - \mye^{-(H-H^*)/(d^*H^*)}},
\end{align}
where $A^*$ is the Hamaker constant of each respective model, $n^*$ and $m^*$ (with $n^*>m^*$) are positive integers characterising the strength of the intermolecular interactions, and $d^*<1/(m^*-1)$ gives the lengthscale associated with the decay of the repulsive force. To find these disjoining potentials in nondimensional form, we use the scaling in \refe{eq:ndscalsdp}, and a version of \refe{eq:earlyangledp}, to give a relationship between the Hamaker constant and the contact angle (which is scaled to unity in our choice of the nondimensionalisation here). Thus we use
\begin{align}
 V(H^*) = -\frac{\sigma}{2}, \qquad \Pi(H) = \pfrac{V}{H},
\end{align}
and find the two forms
\begin{align}
 \Pi(h) &= \frac{(m^*-1)(n^*-1)}{2\epsilon(n^*-m^*)}\lrsq{\lr{\frac{\epsilon}{h}}^{m^*} - \lr{\frac{\epsilon}{h}}^{n^*}}, \nonumber\\
 \Pi(h) &= \frac{m^*-1}{2\epsilon[1-(m^*-1)d^*]}\lrsq{\lr{\frac{\epsilon}{h}}^{m^*} - \mye^{-(h-\epsilon)/(d^*\epsilon)}},
\end{align}
where $\epsilon=H^*/L\ll1$, is the small parameter associated with these disjoining pressure models (analogous to $\lambda$ for the slip models), see \cite{SavvaPrecursorSlip} for further details. The droplet spreading geometry is given in figure \ref{fig:drop}, where the small parameter $\delta$ is used, and corresponds to either $\lambda$ for slip models (when $\epsilon=0$), or $\epsilon$ for disjoining pressure models (when $\lambda=0)$. We reiterate that these disjoining pressure models create a thin precursor film ahead of the contact line and extending to immediately cover the entire substrate (although see discussion by de Gennes about this point \cite{deGennesrev}), so that there is no actual contact line and thus no need for slip to resolve an associated contact line singularity (as \eg \cite{hocking95precursor}). Other, more complex forms of disjoining pressure may be considered where actual contact lines exist and slip is necessary for the motion of contact lines, see e.g.~\cite{WuWongJFM}.

Another model has been proposed based on diffuse-interface theory.
Following \cite{PismenPomeau}, we consider a
liquid-gas diffuse-interface as the \emph{compressible} Navier--Stokes
equations where the stress tensor is given by the compressible Newtonian viscous stress tensor with a diffusion tensor added. This diffusion tensor (also known as the capillary or Korteweg tensor) incorporates terms which are associated with
the pressure and surface tension, arising from the interaction of the molecules within the interface, and is given by
\begin{align}
 \bar{\textbf{T}}_D = \lrsq{K\lrmod{\bar\bmN \bar\rho}^2/2 + K\bar\rho\bar\nabla^2\bar\rho - \bar{p}}\textbf{I} - K\bar\bmN\bar\rho\otimes\bar\bmN\bar\rho,
\end{align}
where $K$ is a gradient energy coefficient.
The fluid density $\bar\rho$
acts as an order parameter that varies between the two equilibrium states of the fluid
(\ie the bulk liquid and vapour densities $\bar\rho_l$ and $\bar\rho_v$
respectively). The precise diffuse-interface governing equations, with no-slip applied on the wall, are given in \cite{myEPJE}. The equations were studied in \cite{myEPJE} without making the long-wave assumption, and it was shown that the singularities associated with the contact line problem are resolved without requiring a precursor film (\ie the diffuse-interface model alone is sufficient mathematically). This analysis was also extended to include general free-energy and viscosity forms in \cite{myDIpof}. However, if modelling a situation where a precursor film is desired, then a boundary condition $\bar{\rho}=\bar\rho_w$ (where $\bar\rho_w$ is a prescribed wall density) may be applied. Pismen and Pomeau \cite{PismenPomeau} considered this case, and found that in the thin-film approximation, the model reduces to the evolution equation \refe{eq:dimlw}, with $\bar\lambda=0$ and disjoining pressure
\begin{align}
 \bar\Pi(H) = 2A^*\lrsq{\lr{\frac{\mye^{-H/H^*}}{w}} - \lr{\frac{\mye^{-H/H^*}}{w}}^{2}},
\end{align}
where $A^*$ is now an effective Hamaker constant, being the strength of the molecular interactions, $H^*$ is strictly now the width of the diffuse interface, rather than the exact thickness of the precursor film, and $w = 1 - \bar\rho_w / \bar\rho_l$, is a small positive parameter being a measure of the density at the wall \ie a measure of the wetting properties of the substrate.
Using the same nondimensionalisation procedure outlined above, we find
\begin{align}
 \Pi(h) = \frac{w^2\mye^2}{\epsilon(2w\mye-1)}\lrsq{\lr{\frac{\mye^{-h/\epsilon}}{w}} - \lr{\frac{\mye^{-h/\epsilon}}{w}}^2}.
\end{align}

\begin{figure}[ht]
\centering
\includegraphics[width=12cm]{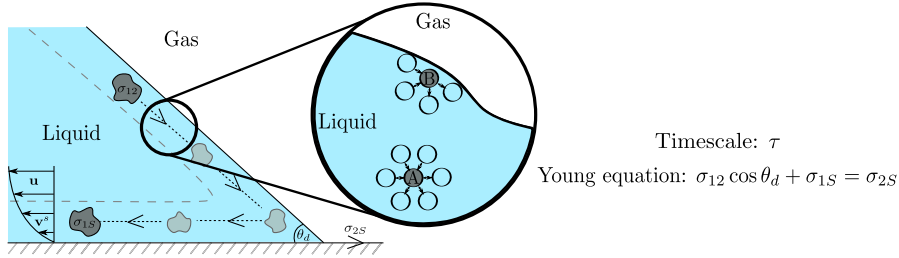}
\caption{Illustrative schematic of some principal features of the interface formation model. The interface at the solid is shown with velocity $\tbV^s$ satisfying no-slip, yet effective slip for bulk velocity $\tbU$. A magnification of the liquid interface displays the density variation close to an interface, resulting in the surface tension---where fluid particle A away from the interface is surrounded by other fluid particles, whereas the fluid particle B near the interface feels the presence of fewer fluid particles and is thus in an unfavourable state. The surface tensions satisfy the Young equation, and vary over a finite timescale $\tau$. This schematic is purely illustrative, as the interfaces are taken with zero thickness in the model.}
\label{fig:ifmschem}
\end{figure}

Finally, we discuss the interface formation model, which was studied in the
thin-film regime for droplet spreading in
\cite{My_Shikh}.
The interface formation model exploits distinct surface variables and dynamic surface tensions and is more closely linked to slip models than diffuse-interface methods, albeit in a rather complex and sophisticated form. It was developed in 1993 by Shikhmurzaev \cite{Shikh93}, (thus predating diffuse-interface approaches for modelling contact lines, starting from the work of Seppecher in 1996,~\cite{Seppecher}) and has been used in a variety of wetting situations. The interface formation model has attracted some controversy, however (see the recent methodical analysis by the present authors on the model utilising the thin-film regime \cite{My_Shikh}, and the monograph of Shikhmurzaev \cite{ShikhBook}). Briefly, the key physical ingredients of the interface formation model are that:
\begin{itemize}
 \item the interfaces between liquid and gas, and liquid and solid, have surface variables distinct from bulk values.
 \item the surface tension varies dynamically, and relaxes on a finite timescale $\tau$. The contact angle is also assumed to vary dynamically (and will be given as a solution of the equations, rather than being prescribed as for the slip models),
\end{itemize}
and a purely illustrative schematic is shown in figure \ref{fig:ifmschem}.

At the contact line the interface formation model satisfies both the Young equation \cite{Young}, and conservation of mass (which, unlike for slip models where this holds automatically, mass conservation at the contact line must be explicitly imposed in the model). The Young equation describes a force balance in the direction parallel to the solid surface. Considering a static contact line region, then the force balance is
\begin{align}
 \sigma_{12}\cos\theta_c + \sigma_{1S} = \sigma_{2S}, \label{eq:youngeq}
\end{align}
where $\sigma_{12}$, $\sigma_{1S}$ and $\sigma_{2S}$, are the surface tensions of the fluid 1-fluid 2, fluid 1-solid and fluid 2-solid interfaces, and $\theta_c$ is the contact angle. For the application of the interface formation model, the Young equation \refe{eq:youngeq} is assumed to hold dynamically (see, for example Sec. 4.3.5.2 of Ref. \cite{ShikhBook}), with the microscopic contact angle $\theta_c$ changing from its static value $\theta_c = \theta_S$ to a dynamically varying angle $\theta_c = \theta_d$. This then requires at least one of the surface tensions to also change from its static value and provides the motivation for, and the distinguishing feature of, the interface formation model. This is as inherent within the interface formation model is that the surface tensions of the interfaces dynamically vary on a finite timescale $\tau$ (as discussed above).

The system of equations to solve is comparatively complex,
being the usual incompressible Navier--Stokes equations in the bulk, but
coupled to a system of six equations forming the boundary conditions at the
solid-liquid interface, seven at the liquid-gas free surface and two further
boundary conditions at the contact line (from the Young equation and from
conservation of mass through the contact line). Due to their abundance and
complexity, we will not give details of the full equations here, but refer
the reader to \cite{My_Shikh} and to the monograph \cite{ShikhBook}. After a process of nondimensionalisation, the thin-film
equations of the interface formation model for the droplet spreading problem
are (see \cite{My_Shikh} for details)
\begin{equation}
 \pfrac{\Bah}{\Bat} + \pfrac{}{\Bax}\left(-\frac{1}{2}\Bah^3 \pppfrac{\Bah}{\Bax}
+ \frac{3}{2}A\Bah^2 + B\Bah\right)
= \bar{R}_0 \mathcal{D} ,\label{eq:ifmlw}
\end{equation}
where $A(x,t)$ and $B(x,t)$ give effectively the shear stress and the slip velocity at the solid surface, respectively, and satisfy the two additional equations
\begin{subequations}
\begin{align}\label{eq:chi0AB1}
\bar\beta B - A &=
\bar\tau\breve{\rho} \ppfrac{}{\Bax}
\left(
\frac{1+4{\bar\alpha}}{4}B -  \frac{\bar\alpha}{\bar\beta}A   \right),
\\
\Bah\pppfrac{\Bah}{\Bax}-A &=
\bar\tau\ppfrac{}{\Bax}
\left[ \frac{3}{2} \Bah^2\pppfrac{\Bah}{\Bax} - 3A\Bah - B
+ \frac{1+4{\bar\alpha}}{4\bar\beta}
\left( \Bah\pppfrac{\Bah}{\Bax}-A \right)
 \right] .\label{eq:chi0AB2}
\end{align}
\end{subequations}
These three governing equations contain the nondimensional parameters
$\bar{R}_0$, $\bar\beta$, $\bar\tau$, $\breve{\rho}$, and $\bar\alpha$, which
correspond to nondimensional numbers characterising the magnitude of mass
transfer between bulk and surfaces, an inverse slip coefficient, the
nondimensional surface tension relaxation time, the ratio of the equilibrium
solid and gas surface densities, and a nondimensional number combining
Darcy-type and slip length parameters to characterise the surfaces respectively---further
details again in \cite{My_Shikh}. Finally, $\mathcal{D}$
represents the sum of the dynamic surface densities and is given by
\begin{equation}
\mathcal{D} = {\bar\tau}\pfrac{}{\Bax} \left[
\frac{1+4{\bar\alpha}}{ 4\bar\beta} \left( \Bah\pppfrac{\Bah}{\Bax}-A
- 2{\breve{\rho}}\bar\beta B
 \right)
+
   \frac{2{\breve{\rho}}{\bar\alpha}}{\bar\beta}A    +\frac{3}{2}\Bah^2
\pppfrac{\Bah}{\Bax}  - 3A\Bah - B\right].\label{eq:chi0rhoSG}
\end{equation}
These long-wave equations are substantially more complex than their slip or precursor film counterparts, but if $\bar\tau=0$ so that the surface tension relaxes instantaneously---\ie the surface tensions will be constant, then the Navier--slip model is recovered. At the contact line ($\Bax=a$), the two boundary conditions imposed, related to the mass balance and the Young equation are
\begin{align}
(1 + 4\bar\alpha)\lrsq{\frac{A}{4\bar\beta} + \frac{\breve{\rho}B}{2} - \frac{\Bah}{4\bar\beta}\pppfrac{\Bah}{\Bax}}
- \frac{2\breve{\rho}\bar\alpha A}{\bar\beta} + B = \bar{U}_{CL}( 1+ \breve{\rho}),
\quad \mbox{and} \quad
 1- \lrsq{\pfrac{\Bah}{\Bax}}^2 = 2\mathcal{D} ,
\end{align}
where $ \bar{U}_{CL} = B - \bar R_0\left(\partial_{\Bax}\Bah\right)^{-1}\mathcal{D} $.

\subsection{Quasistatic spreading}

We now consider the prototype situation of quasistatic spreading of a thin two-dimensional droplet, with the geometry as shown in figure \ref{fig:drop}. The evolution of the droplet radius $a(t)$ will be determined through the method of matched asymptotics. In the bulk of the droplet (the outer region), the leading order behaviour is modelled with the thin-film evolution equation with negligible slip, disjoining pressure or other microscopic effects, whereas these play a significant role in the inner regions near the contact lines. The outer region behaviour holds for all models considered in this work, \ie for the slip models, the disjoining pressure models and the interface formation model, and arises from neglecting the microscopic effects in \refe{eq:ndlw} or \refe{eq:ifmlw}--\refe{eq:chi0rhoSG}. The inner region behaviours for the variety of slip models will then be considered in detail, with the remaining models then discussed and compared at the end of this section.

\paragraph*{Outer region:}

The microscopic effects of slip, disjoining pressure etc., are not significant in the outer region, and as such we take the limit $\delta\to0$ (and equivalently $\bar \beta\to \infinity$, with $\bar \beta\bar\tau=O(1)$, for the interface formation model). Subsequent calculations are now given at leading order in an asymptotic expansion in $\delta$.  We transform the symmetric droplet domain $-a(t)\leq\Bax\leq a(t)$ to $-1\leq y \leq 1$ via the mapping $\Bax=ay$ so that the governing equation in terms of the (non-moving) coordinate $y$ becomes
 \begin{align}\label{eq:actualprob}
 \pfrac{\Bah}{\Bat} - \frac{\dot{a}y}{a}\pfrac{\Bah}{y} +
 \frac{1}{a^4}\pfrac{}{y}\left[
\Bah^3 \pppfrac{\Bah}{y}\right] = 0,
\end{align}
where $\dot{a}=da/dt$ is the spreading rate of the contact line. Here, we are considering slow spreading, and thus
consider the quasistatic limit $|\dot{a}|\ll1$ with expansions of the form
\begin{align}
 h(y,t) = h_0(y,a) + h_1(y,a,\dot{a}) + \cdots,
\label{eq:outerexpan}
\end{align}
where the time dependence of $h$ enters through $a$ and its time derivatives.
An assumption made in this procedure is that $\delta\ll|\dot{a}|$. This
will be found to be consistent as $\dot{a}$ depends only logarithmically on
$\delta$ [see later in eq.~\refe{eq:srr}].

At leading order in our quasistatic expansion we have
 \begin{equation}
\pfrac{}{y}\left( \Bah_0^3\pppfrac{\Bah_0}{y} \right)=0,
\qquad \mbox{with conditions} \qquad
  \Bah_0(1) = 0, \quad \left.\pfrac{\Bah_0}{y}\right|_{y=0}=0,
\quad \int_0^1 \Bah_0 dy = \frac{1}{a}. \label{eq:outle}
\end{equation}
These conditions correspond to zero height at the contact line, a symmetry condition such that the slope of the droplet is zero at its centre, and a normalization condition---such that the nondimensional cross sectional area of the half-droplet is unity. Equations \refe{eq:outle} are solved to find
\begin{equation}
 \Bah_0 = \frac{3(1-y^2)}{2a}. \label{eq:H0sol}
\end{equation}
The equation for the next term is
\begin{equation}
\pppfrac{\Bah_1}{y} =  \frac{4\dot{a} a^5 y}{ 9(1-y^2)^2},
\end{equation}
giving
\begin{equation}
\Bah_1 = \frac{2}{18}\dot{a}a^5\lrsq{ y\ln\lr{\frac{1+y}{1-y}} + \ln[(1+y)(1-y)]-\frac{3y^2}{2}
+\frac{1}{2}\ln\lr{\frac{\mye^3}{16}}}
,
\end{equation}
having used $\pilfrac{\Bah_0}{\Bat}=\dot{a}\pilfrac{\Bah_0}{a}$, and where arbitrary constants have been determined from $h_1(1)=0$, $(\pilfrac{\Bah_1}{y})(0)=0$, and $\int_0^1 \Bah_1 dy = 0$. Using these results for $h_0$ and $h_1$, and transforming back to the original coordinate $x$, the
behaviour of the free surface slope in the outer region as the inner region is approached (as $\Bax\to a$) is found as
\begin{equation}
 -\pfrac{\Bah}{\Bax} \sim \frac{3}{a^2} + \frac{1}{9}{\dot{a}a^4}\ln\left[\frac{\mye^3(a-\Bax)}{2a}\right].\label{eq:outbehav}
\end{equation}
A similar behaviour is found for $\Bax\to-a$ for the left contact point, but the droplet symmetry allows us to consider only the right contact point in the remaining analysis.

\paragraph*{Inner region:}
The behaviour from \refe{eq:outbehav} must be matched to its inner region counterpart, where the microscopic effects are significant (and as discussed only the inner region near $\Bax=a$ being considered due to the droplet symmetry). The inner region is defined by the change of variables
\begin{align}
 h=\delta\Phi, \qquad x = a-\delta\xi,
\end{align}
which, for the slip models, yields the inner region evolution equation
\begin{align}
 \delta\pfrac{\Phi}{t} + \dot{a}\pfrac{\Phi}{\xi} + \pfrac{}{\xi}\lrcur{
 \Phi^2 \lr{\Phi + \frac{1}{\Phi^n}}
 \pppfrac{\Phi}{\xi} }=0. \label{eq:ndlwin}
\end{align}
Proceeding with a quasistatic expansion  $\Phi(\xi,t) = \Phi_0(\xi,a) + \Phi_1(\xi,a,\dot{a}) + \cdots$, then at leading order
\begin{align}
 \lr{\Phi_0^3 + \Phi_0^{2-n}} \pppfrac{\Phi_0}{\xi} =\mbox{Const}_1.
\end{align}
For $n=\{0,1\}$, corresponding to Navier-slip and inverse linear slip, the integration constant must be zero to satisfy the boundary condition of zero height at the contact line $\Phi|_{\xi=0}=0$. In these cases, the nontrivial solution is given by $\pppilfrac{\Phi_0}{\xi}=0$, and on solving this subject to $\Phi_0|_{\xi=0}=0$ and $\Phi_0\xi^{-2}\to0$ as $\xi\to\infinity$ (to be able to match to the outer region $\xi\ln\xi$ behaviour), then $\Phi_0 = C_a \xi$, with the arbitrary constant $C_a$ being fixed by the contact angle condition. For $n=2$ there will  be other possible solutions to this leading order equation $\pppilfrac{\Phi_0}{\xi} = [\pppilfrac{\Phi_0}{\xi}|_{\xi=0}]/(\Phi_0^3 + 1)$, but the wedge solution $\Phi_0 = C_a \xi$ holds. As it would be expected that the leading order solution is unaffected by the slip model chosen, we assume this wedge solution is the physically relevant solution for the $n=2$ case also.

At the next order, the equation to solve (after integrating once) is then
\begin{align}
 C_a\dot{a}\xi + (C_a^3\xi^3 + C_a^{2-n}\xi^{2-n})\pppfrac{\Phi_1}{\xi} = \mbox{Const}_2,
\end{align}
and once again the integration constant must be zero for $n=\{0,1\}$, but satisfies
$\mbox{Const}_2 = \pppilfrac{\Phi_1}{\xi}|_{\xi=0}$ for $n=2$. It is found, however, that $\pppilfrac{\Phi_1}{\xi}|_{\xi=0}$ must vanish to prevent a singular velocity at the contact line for the $n=2$ slip model. For $n=0$ then, the Navier-slip case, the solution satisfying $\Phi_1|_{\xi=0}=0$ and $\Phi_1\xi^{-2}\to0$ as $\xi\to\infinity$ is
\begin{align}
 \Phi_1 =
 \frac{\dot{a}}{2C_a^3}\lrsq{(C_a\xi+1)^2\ln(C_a\xi+1)-(C_a\xi \ln(C_a\xi)-2 C_b C_a^2+1)C_a\xi}, \label{eq:p1_0}
\end{align}
equivalently for $n=1$, inverse linear slip, it is
\begin{align}
 \Phi_1 =
 \frac{\dot{a}}{2C_a^3}\lrsq{C_a\xi\ln(1+C_a^2\xi^2)+(1-C_a^2\xi^2)\arctan(C_a\xi)+\frac{\xi}{2}(C_a^2\pi\xi-2C_a+4C_bC_a^3)},
  \label{eq:p1_1}
\end{align}
and for $n=2$, the solution is
\begin{align}
 \Phi_1 &=
 \frac{\dot{a}}{12C_a^3}\left[
 2\sqrt 3(1-C_a^2\xi^2) \arctan\lrcur{(2C_a\xi-1)/\sqrt 3}\right.\nonumber\\&\left.
 - (1+C_a^2\xi^2 - 4C_a\xi)\ln(C_a^2\xi^2-C_a\xi+1)
 + 2(C_a\xi+1)^2\ln(C_a\xi+1)\right.\nonumber\\&\left.
 + \pi\sqrt 3(C_a^2\xi^2+{3}^{-1}) - 6C_a(1-2C_a^2C_b)\xi
 \right], \label{eq:p1_2}
\end{align}
where in all three cases, another arbitrary constant $C_b$ will be fixed by the contact angle condition.

Alongside the crucial behaviour as $\xi\to\infinity$ to match to the outer region, there are three other important quantities to consider at the contact line $\xi=0$, these being given in original nondimensional variables in \refe{eq:nd_pvs}, and in inner variables (with $p=\delta^{-1}p_i$, $u=u_i$, and $z=\delta z_i$) becoming
\begin{gather}
 p_i = -3\ppfrac{\Phi}{\xi} = -3\ppfrac{\Phi_1}{\xi} + \cdots, \qquad
 u_i|_{z_i=0} = -\Phi^{1-n}\pppfrac{\Phi}{\xi} = -(C_a\xi)^{1-n}\pppfrac{\Phi_1}{\xi} + \cdots, \nonumber\\
 \lreval{\pfrac{u_i}{z_i}}_{z_i=0} = -3 C_a\xi \pppfrac{\Phi_1}{\xi} + \cdots.
\end{gather}
Using these we find
\begin{align}
 n=0: \quad& p_i = \frac{3\dot{a}}{C_a}[\ln(\xi)+\ln(C_a)] + O(\xi), \quad
 u_i|_{z_i=0} = \dot{a} + O(\xi), \quad
 \lreval{\pfrac{u_i}{z_i}}_{z_i=0} = 3 \dot{a} + O(\xi), \nonumber\\
 n=1: \quad& p_i = -\frac{3\dot{a}\pi}{2C_a} + O(\xi), \quad
 u_i|_{z_i=0} = \dot{a} + O(\xi^2), \quad
 \lreval{\pfrac{u_i}{z_i}}_{z_i=0} = 3 C_a \dot{a} \xi + O(\xi^3), \nonumber\\
 n=2: \quad& p_i = -\frac{2\pi\dot{a}}{\sqrt{3}C_a} + O(\xi^2), \quad
 u_i|_{z_i=0} = \dot{a} + O(\xi^3), \quad
 \lreval{\pfrac{u_i}{z_i}}_{z_i=0} = 3 C_a^2 \dot{a} \xi^2 + O(\xi^5),
\end{align}
where we see that the stress is regularised by all slip models, the velocity at the contact line is always $\dot{a}$ as expected, and the logarithmic pressure singularity for Navier-slip persists in this prototype system. We note that for the cases $n=\{1,2\}$, there is a constant term in the pressure due to the non-zero curvature of the free-surface at this order. It is also of interest to compare to the Moffatt wedge flow solutions, given in Appendix~\ref{app:wedge}, eqs. \refe{appeq:ns}--\refe{eq:tocomp_p_thin}. Accounting for the change of reference frame and the previous remark about non-zero curvature, we see agreement between order of magnitudes for all of these variables, noting that $r\sim\xi$ to compare between Moffatt wedge and thin droplet settings.

For the behaviour of the free-surface slope to match to the outer region, all three slip models have the \emph{same} behaviour at leading order, that of
\begin{align}
 \pfrac{\Phi_1}{\xi} \sim \dot{a}\lrsq{\frac{\ln\xi}{C_a^2} + \frac{1+\ln C_a}{C_a^2} + C_b}, \quad \mbox{i.e.} \quad
\pfrac{\Phi}{\xi} \sim C_a + \dot{a}\lrsq{\frac{\ln\xi}{C_a^2} + \frac{1+\ln C_a}{C_a^2} + C_b},
\quad \mbox{as }\xi\to\infinity,
\label{eq:innmatch}
\end{align}
and thus
\begin{align}
 -\pfrac{h}{x} \sim C_a + \dot{a}\lrsq{\frac{\ln\lr{(a-x)/\delta}}{C_a^2} + \frac{1+\ln C_a}{C_a^2} + C_b}, \quad
 \mbox{as }(a-x)/\delta\to\infinity,\label{eq:innmatchh}
\end{align}
and, as performed in \cite{Savva09}, matching the cube of the free-surface slope yields the evolution equation
\begin{align}
 3\dot{a}a^6\ln\lrsq{\frac{\delta \, \mye^{2 - C_b C_a^2}}{2 a C_a}} = C_a^3 a^6 - 27, \label{eq:srr}
\end{align}
with $C_a$ and $C_b$ still to be set by the imposed contact angle condition, which we will now consider. The matching of the cube of the free surface slope may be justified through asymptotics, and is due to the specific structure of expressions \refe{eq:outbehav} and \refe{eq:innmatchh}---where for the logarithmic terms in $(a-x)$ to balance, a cubic power is necessary. Motivated by the asymptotic analysis of Hocking, \cite{Hocking83}, where matching was performed between inner and outer regions through another intermediate region (the purpose of the intermediate region purely to justify the matching of the cubes \cite{Savva09}), we describe an intermediate region relevant for our analysis in Appendix~\ref{app:inter}, for the interested reader.

\paragraph*{Contact angle condition for slip models:}

As discussed in the introduction, a number of different contact angle conditions have been imposed in the literature for slip models. The conditions discussed were that the dynamic contact angle always equals the static value, and two other relations in \refe{eq:introCArel}, and in the thin-film regime here these correspond (for the right contact point) to
\begin{align}
 \mbox{(i): } \pilfrac{h}{x} = -1, \qquad
 \mbox{(ii): } \dot{a} = k(-\pilfrac{h}{x}-1)^m, \qquad
 \mbox{and} \qquad
 \mbox{(iii): } \dot{a} = k([-\pilfrac{h}{x}]^m-1).
\end{align}
Contact angle condition (i) for thin droplet spreading has been used in \cite{Hocking83,Savva09,SavvaPrecursorSlip}. In this instance the asymptotic procedure follows as above, with, in the inner region $\pilfrac{\Phi}{\xi}(0)=\pilfrac{\Phi_0}{\xi}(0)+\pilfrac{\Phi_1}{\xi}(0)+\cdots=C_a+\pilfrac{\Phi_1}{\xi}(0)+\cdots=1$. This thus determines $C_a=1$, and on using the solutions from \refe{eq:p1_0}--\refe{eq:p1_2}, we find $\pilfrac{\Phi_1}{\xi}(0)=C_b\dot{a}$, which must vanish for this boundary condition and hence $C_b=0$.

In contrast, both contact angle conditions (ii) and (iii) can give a leading order solution for the spreading rate using only the outer solution, provided $k=O(\dot{a})$, as shown by Hocking \cite{HockingRival} for a form of condition (ii) and based on work in \cite{greenspan,HaleyMiksis,ehrharddavis}. In this case the evolution is of the form
$\dot{a} = k(3/a^2 - 1)^m$ for (ii) and
$\dot{a} = k([3/a^2]^m - 1)$
for (iii), forming very different evolutions\footnote{It should be noted here that whilst the contact angle condition can be applied directly to the behaviour of the outer solution as the contact line is approached, a slip condition at the wall is still required to alleviate the stress singularity, and inner regions are necessary to compute the correction to the leading order in the evolution equation for the droplet radius. See \cite{HockingRival} for further details.}. Such an imposed order-of-magnitude for $k$, ostensibly just a constant of proportionality (and thus perhaps expected to be $O(1)$), appears rather restrictive. Additionally, if no inner (slip) region is sought at leading order for the droplet evolution, care should be taken to ensure that conservation of mass for the droplet is not violated.

If the asymptotic analysis as performed for contact angle condition (i) is considered, then contact angle conditions (ii) and (iii) in the inner region give respectively
\begin{align}
 (\dot{a}/k)^{1/m} = C_a - 1 + \pilfrac{\Phi_1}{\xi}(0) + \cdots, \qquad
 \dot{a}/k = C_a^m - 1 + m C_a^{m-1} \pilfrac{\Phi_1}{\xi}(0) + \cdots,
\end{align}
and thus provided $\dot{a}/k=o(1)$, which is in particular true if $k=O(1)$, then $C_a=1$. As $\Phi_1=O(\dot{a})$, then contact angle condition (ii) is only valid in this regime if $(\dot{a}/k)^{1/m} = O(\dot{a})$, \ie if $k = O(\dot{a}^{1-m})$, whereas (iii) is valid for any $k$ satisfying $k \gg \dot{a} $ (due to the requirement for $C_a=1$). As $\pilfrac{\Phi_1}{\xi}(0)=C_b\dot{a}$ for the slip models in \refe{eq:p1_0}--\refe{eq:p1_2}, then if $k \gg \dot{a}^0$ in (iii)
then $C_b=0$ as for contact angle condition (i). Otherwise, for conditions (ii) and (iii) respectively
\begin{align}
 C_b = \frac{\dot{a}^{(1-m)/m}}{k^{1/m}}, \qquad C_b = \frac{1}{m k}.\label{eq:Cbvals}
\end{align}
We reiterate that if $k$, which originally appeared as a constant of proportionality and most naturally would not be expected to depend on the spreading rate, is $O(1)$, then this asymptotic analysis fails for boundary condition (ii) unless $m=1$. In this case, both conditions (ii) and (iii) are equivalent, and so for $k=O(1)$, boundary condition (iii) is preferred.

The slip models satisfy the spreading rate equation
\begin{align}
  3\dot{a}a^6\ln\lrsq{\ilfrac{\delta \, \mye^{2 - C_b}}{(2 a)}} = a^6 - 27.\label{eq:sr_slip}
\end{align}
We finally note that provided $a(t)$ is far from the equilibrium value, $a(\infinity)=\sqrt{3}$, equation \refe{eq:sr_slip} gives the radius behaviour of $a=(ct)^{1/7}$, where $c=-63/\ln[\ilfrac{\delta \, \mye^{2 - C_b}}{(2 a)}]$. This logarithmic correction to a $t^{1/7}$ behaviour agrees with the rigorous result of \cite{GiacomelliOtto}, who showed that using a slip model for droplet spreading alters the naively expected spreading rate only logarithmically through deriving suitable estimates for integral quantities such as the free energy and the length of apparent support.

\paragraph*{Precursor film models:}

For the precursor film models, we follow \cite{SavvaPrecursorSlip,PismenEggersSolv} and find the inner region behaviour when the outer region is approached (and in outer variables) as
\begin{align}
 -\pfrac{h}{x} \sim 1 + \dot{a}\ln\lr{\frac{\mye (a-x)}{\ell\delta}}, \quad
 \mbox{as }(a-x)/\delta\to\infinity, \label{eq:pfinbehav}
\end{align}
where
\begin{align}
\ln\ell = \int_1^\infinity \lr{\frac{1}{\zeta}-\frac{(\zeta-1)^2}{\zeta^3\sqrt{1+K(\zeta)}}}\ \textrm{d}\zeta, \label{eq:ell}
\end{align}
and $K(\zeta) = 2\epsilon\int \Pi(\epsilon \zeta) \ \textrm{d}\zeta$,
which for the three disjoining pressure models (Schwartz and Eley, Sharma, and Pismen and Pomeau, respectively) gives
\begin{gather}
 K(\zeta) = \frac{1}{n^*-m^*}\lr{\frac{m^*-1}{\zeta^{n^*-1}}-\frac{n^*-1}{\zeta^{m^*-1}}}, \qquad
 K(\zeta) = \frac{d(m^*-1)\mye^{(1-\zeta)/d}-{\zeta^{1-m^*}}}{1-d(m^*-1)}, \nonumber\\
 K(\zeta) = -\mye^{1-\zeta} - \frac{1}{2w\mye-1}\lr{\mye^{1-\zeta}-\mye^{2(1-\zeta)}}.
\end{gather}
Finally, matching the behaviour from \refe{eq:pfinbehav} to the outer behaviour gives
\begin{align}
 3\dot{a}a^6\ln\lrsq{\ilfrac{\mye^2 \, \delta\ell}{(2 a)}} = a^6 - 27. \label{eq:sr_pf}
\end{align}

\paragraph*{Comparisons:}

The last model we will discuss in this section is the interface formation model, whose analysis is substantially more involved. However, it was shown in \cite{My_Shikh} that the equivalent spreading rate is given by
\begin{align}
 3\dot{a}a^6\ln\lrsq{\ilfrac{\mye^{2-c_0}}{(2 a \bar\beta)}} = a^6 - 27, \label{eq:sr_ifm}
\end{align}
where $c_0$ was numerically determined from a seventh order system of ODEs, dependent upon the nondimensional numbers associated with the model---$\bar{R}_0$, $\bar\beta$, $\bar\tau$, $\breve{\rho}$, and $\bar\alpha$. We are now able to draw comparisons between all of these models in this droplet spreading situation.

As we saw in \refe{eq:sr_slip}, the three slip models (\ie Navier--slip, and inverse linear/quadratic slip) all give equivalent spreading rates, but their behaviour depends on the contact angle condition through $C_b$, which is zero for condition (i), but nonzero and given in \refe{eq:Cbvals} for (ii) and (iii), provided the coefficient of proportionality is in the regime considered.

The spreading rates obtained through matching for these slip models become equal if the slip length for condition (i), $\lambda=\lambda_{(i)}$ is given by
$\lambda_{(i)} = \lambda_{(*)}\mye^{-C_b}$, where $\lambda_{(*)}=\{\lambda_{ii},\lambda_{iii}\}$. The slip and precursor film models give equivalent spreading rates if $\lambda\mye^{-C_b} = \epsilon \ell$ (seen from \refe{eq:sr_pf}), and the individual precursor film models comparable, but reliant on numerically computing the integrals in \refe{eq:ell}.

Finally, these models predict equivalent spreading rates to the interface formation model (seen in \refe{eq:sr_ifm}) if $\lambda\mye^{c_0-C_b} = \bar\beta^{-1}$, or $\mye^{c_0} \epsilon \ell = \bar\beta^{-1}$. Thus, even though the original PDEs for these models are different, we obtain nearly identical behaviours if the parameters (\eg parameters $\lambda$, $C_b$, $\epsilon$, and $\bar\beta$) are chosen appropriately. These results are summarised in table \ref{tab}. As these results are for the prototype slow spreading thin droplet situation, it would be of significant interest to see comparisons between the various models in situations where the underlying physics they capture is likely to be significant, such as for rapid wetting where dynamic contact angle variation could be more pronounced (although noting that contact angles must remain small for the thin-film approximation to remain valid). The work of Eggers \cite{eggershigherCa}, which extends the analysis beyond the quasistatic assumption for Navier-slip and inverse linear slip models with BC (i), provides some results of relevance in this pursuit.

\begin{table}
\caption{Comparison of the models, showing what the microscopic parameter for
each model should be to give an equivalent spreading rate to the interface
formation model. This comparison is for quasistatic thin two-dimensional
droplet spreading, however the equivalence is also expected to persist in 3D
spreading provided that the contact line variations occur at much longer
lengthscales compared to the microscopic parameters. Note that the
Navier-slip, and inverse linear/quadratic slip are all called `slip' as they
all have equivalent spreading rate expressions.}\label{tab}
\begin{tabular}{cccc} \hline
Model & $\begin{array}{c}
\mbox{Condition at the}\\\mbox{contact line (for slip)}
\end{array}$ & $\begin{array}{c}
\mbox{Micro-}\\\mbox{scopic}\\\mbox{parameter}
\end{array}$ & $\begin{array}{c}
\mbox{Choice of micro-}\\\mbox{scopic parameter for}\\\mbox{equivalent spreading }
\end{array}$\\ \hline
Interface formation & \rule[0.5ex]{2em}{0.55pt} & $\bar\beta^{-1}$ & \rule[0.5ex]{2em}{0.55pt}\\
Slip with BC (i) &  $\pilfrac{h}{x} = -1$ & $\lambda$ & $\bar\beta^{-1}\mye^{-c_0}$\\
Slip with BC (ii) &  $\dot{a} = k(-\pilfrac{h}{x}-1)^m$ & $\lambda$ & $\bar\beta^{-1}\mye^{C_b-c_0}$, $C_b$ in \refe{eq:Cbvals}(a)\\
Slip with BC (iii) &  $\dot{a} = k([-\pilfrac{h}{x}]^m-1)$ & $\lambda$ & $\bar\beta^{-1}\mye^{C_b-c_0}$, $C_b$ in \refe{eq:Cbvals}(b)\\
Precursor film & \rule[0.5ex]{2em}{0.55pt} & $\epsilon$ & $\bar\beta^{-1}\mye^{-c_0}\ell^{-1}$\\
\hline
\end{tabular}
\end{table}

\section{Conclusions}
We have revisited the problem of thin two-dimensional droplet spreading, a useful prototype system involving moving contact lines, and extended the comparative analysis of Savva and Kalliadasis \cite{SavvaPrecursorSlip} to include a disjoining pressure model arising from diffuse-interface theory, a variety of contact angle conditions applied to slip models
[specifically the extension to give the results for slip with BC (ii) and (iii) in table~\ref{tab}],
and compared to results from Sibley \etal~on the interface formation model
\cite{My_Shikh} in this setting. A summary of the results may be found in
table~\ref{tab}, where an effective redefining of the microscopic parameter
of each model is shown, which will lead to each model giving the same
spreading behaviour as the interface formation model. It should be
stressed that this work is for the slow spreading of thin two-dimensional
droplets. There remains debate in the literature about the best models to
apply when moving contact lines are present, and further studies in other
situations such as rapid/inertial motion
\cite{CoxHighRe,eggershigherCa,FuentesCerroNewModel,SuiJFM13} are needed to
clarify several open questions. Additionally, although there exists a wide
range of models (many of which have been discussed here) there also remains
the need for improved models based on first principles. For example, progress
has been made to more accurately capture effects such as long-range
fluid-fluid interactions \cite{PismenPRE01,Pismenmeso} from the microscale,
and density functional theory appears a promising route
\cite{EvansReview,antoniojfm,BenPRL,BenJPCM}, but the determination of
accurate coarse-grained approximations and associate numerical computations
for contact lines---especially in dynamic settings---is ongoing.

This work compares three different slip models \ie Navier-slip and two nonlinear slip models, to three disjoining pressure models and the interface formation model. The three slip models are based on those of Navier \cite{Navier} for Navier-slip, Ruckenstein and Dunn \cite{RuckDunn}, and Greenspan \cite{greenspan}, for `inverse linear' slip and from Haley and Miksis \cite{HaleyMiksis} for `inverse quadratic' slip. For the prototype quasistatic droplet spreading situation we have investigated it was found that all of these slip models give the same equation for the rate of change of the droplet radius at leading order in the matched asymptotic analysis. As such, all three slip models are described as `slip' in the summary table \ref{tab}.
Three boundary conditions for the contact angle are considered for these three slip models, being that the dynamic contact angle remains at its static value, and two conditions that perturb the dynamic contact angle from its static value through relationships with the contact line velocity. The specific forms can be found in the second column of table \ref{tab}. The effect of including a relationship between dynamic contact angle and contact line velocity is to have an effective shift of the slip-length in the slip models, this shift being related to the specific form of the relationship and its associated proportionality constant and exponent. It is interesting to note that dynamic variation of the contact angle has also been considered in the previous numerical simulations of \cite{ShengZhou92,SuiJFM13} (of a form similar to BC (ii) in table~\ref{tab} here). These simulations also saw an effective shift in slip-length---suggesting the present findings may be robust beyond the thin-film limit.

The three disjoining pressure models are based respectively on van der Waals interactions, a model introduced for polar fluids, and a liquid-gas diffuse-interface model. These three models are used in the works of Schwartz and Eley \cite{SchwartzEley}, Sharma \cite{Sharma}, and Pismen and Pomeau \cite{PismenPomeau}, and all model the situation of an infinite precursor film extending away from the droplet. Once again, these three models all lead to the same equation for the rate of change of the droplet radius in the matched asymptotic analysis, and hence are listed together as `precursor film models' in table \ref{tab}.

We note that these direct comparisons have been made for droplet spreading on a homogeneous substrate. In the previous analysis \cite{SavvaPrecursorSlip}, it was shown that whilst the spreading rate equations from matching may agree, different dynamic behaviour may be realised for spreading on heterogeneous substrates. This is due to the quite different forms of the PDEs, and indeed the physical basis of the models. Unlike for homogeneous substrates, for heterogeneous substrates the system may have multiple equilibria, so that any variation in the dynamics between models will occur most frequently for initial conditions that lie near boundaries separating the basis of attraction to these different equilibria.

The work of \cite{SavvaPrecursorSlip} also included the effects of gravity and of substrate roughness, as mentioned, based on the works in \cite{Savva09,SavvaGrav}. Other features for droplet spreading include chemically heterogeneous substrates \cite{SchwartzEley,Herde12,SchmuckProcA,RajChemHet}, patterned substrates~\cite{DupuisYeomans,KusumaatmajaYeomans}, random (rather than deterministic) substrates \cite{SavvaPavliotisKalliadasis1,SavvaPavliotisKalliadasis2,christophe_marc}, and inclined substrates \cite{ThieleDF2,SavvaInclined}. The understanding of heterogeneous substrates was also considered an important future issue for the contact line problem in the recent review \cite{SnoeijerRev}, and to complement the body of work outlined above on chemical and topological substrate patterns, random disorders and inclinations, it would be of interest to further these analyses by considering how they are affected by the use of the dynamic contact angle models involving slip or interface formation discussed in the present communication.

Finally, we note that the main results here were for two-dimensional droplets, however the main result of the comparability of the models should hold in three dimensions, as well as for other geometries. This is as the asymptotic result relies on the behaviour in the inner asymptotic regions, which in 3D will be the same as for 2D, provided that the contact line variations occur at lengthscales longer than the microscopic lengthscale (\ie $\lambda$ or $\epsilon$).

\section*{Acknowledgements}
We are grateful to the editor and all anonymous referees for the many useful comments and suggestions. We acknowledge financial support from ERC Advanced Grant No. 247031, and Imperial College London through a DTG International Studentship.

%
%

\appendix

\section{Solutions for slip models in the Moffatt wedge geometry}
\label{app:wedge}

The stream function, velocity, stress and pressure behaviours for the Moffatt wedge geometry, for Navier-slip in \refe{eq:psiforms}, are found as
\begin{gather}
 \psi \sim -\frac{r^2 U}{4\lambda}\left( 1-\frac{\theta}{\theta_c} + \frac{\sin(2\theta)\cos(2\theta_c)}{\sin(2\theta_c)}-\cos(2\theta) \right),\nonumber\\
 u \sim -\frac{rU}{4\lambda}\left( -\frac{1}{\theta_c} + \frac{2\cos(2\theta)\cos(2\theta_c)}{\sin(2\theta_c)}+2\sin(2\theta) \right),\qquad
 v \sim -\frac{2\psi}{r}, \qquad p \sim \frac{U}{\lambda \theta_c} \log r,\nonumber\\
 T_{rr} \sim \frac{U}{2\lambda}\frac{\sin(2\theta_c)-2\theta_c\cos(2(\theta_c-\theta))}{\theta_c\sin(2\theta_c)} ,\qquad
 T_{r\theta} \sim \frac{U}{\lambda}\frac{\sin(2(\theta-\theta_c))}{\sin(2\theta_c)} ,\qquad
 T_{\theta\theta} \sim -T_{rr}.\label{appeq:ns}
\end{gather}
These results show clearly how the Navier-slip model removes the singularity in the stress as the contact line is approached, but a logarithmic singularity in the pressure remains.

Similarly for the general slip case in \refe{eq:psiforms}, we obtain
\begin{align}
& \psi \sim r^{2+n}\left( C_{1s}\sin(n\theta)+C_{2s}(\cos(n\theta)-\cos((2+n)\theta)) + C_{3s}\sin((2+n)\theta) \right),
\nonumber\\
& u \sim r^{1+n}\left(n[C_{1s}\cos(n\theta)-C_{2s}\sin(n\theta)]
+(2+n)[C_{3s}\cos((2+n)\theta) + C_{2s}\sin((2+n)\theta)]\right),
\nonumber\\
& v \sim
-\frac{(2+n)\psi}{r}
,\qquad
p \sim 4(1+n) r^n C_{1s} \frac{\cos(n(\theta_c-\theta))}{\cos(n\theta_c)},
\qquad
T_{\theta\theta} \sim -T_{rr},
\nonumber\\
& T_{rr} \sim
2(1+n) r^n \left(
(2+n)[C_{3s}\cos((2+n)\theta) + C_{2s}\sin((2+n)\theta)] + \frac{C_{1s}n\cos(n(\theta_c-\theta))}{\cos(n\theta_c)}
\right)
,\nonumber\\
& T_{r\theta} \sim
2(1+n) r^n \lrsq{C_{2s}\frac{(2+n)\sin((2+n)(\theta_c-\theta))}
{\sin((2+n)\theta_c)}
-C_{1s}
\frac{n\sin(n(\theta-\theta_c))}{\cos(n\theta_c)}
},
\label{eq:tocomp_p_thin}
\end{align}
where
\begin{gather}
 C_{1s} = \frac{U\lambda^{-(1+n)}\sin^n\theta_c}{4(1+n)\tan(n\theta_c)},\qquad
 C_{2s} = - C_{4s} = -C_{1s}\tan(n\theta_c)\qquad
 C_{3s} = \frac{C_{2s}}{\tan((2+n)\theta_c)}.
\end{gather}
We see from this solution that $n=0$ does not recover the Navier-slip solution as this is a degenerate solution of the biharmonic equation. However, the two cases considered by Haley and Miksis \cite{HaleyMiksis}, corresponding to $n=1$ and $n=2$ here, will resolve both the pressure and stress singularities. In contrast to the Navier-slip model, they both give zero pressure and stresses as the contact line is approached---the pressure and stresses being $O(r)$ and $O(r^2)$ in the two cases respectively.

There is, however a possibility of removing the logarithmic pressure singularity but retaining an $O(1)$ stress behaviour. Previously in \refe{eq:presordwlog} we suggested that for the stresses to be $O(1)$ then the stream function must be prescribed by $\psi\sim r^2 F(\theta)$ with $F(\theta)$ to be determined. We have also seen that this form, which corresponds to the Navier-slip condition, yields $O(\log r)$ pressure as $r\to0$ due to the $1/r$ terms in the divergence of the stress. However, it is also possible to cancel the logarithmic pressure terms by relaxing the assumption of constant surface tension, thus obtaining a finite pressure solution. More specifically, from \refe{eq:bctanst}, where $\nabS\sigma$ is the surface gradient of $\sigma$, and which given in terms of the usual gradient operator is $\nabS\sigma = (\iT-\tbN\otimes\tbN)\bmN\sigma$,
we can determine the dynamic boundary condition as
\begin{align}
 T_{r\theta} = -\pfrac{\sigma}{r},\qquad \mbox{on}\quad \theta=\theta_c.\label{eq:bctsgradsig}
\end{align}
Now using the pressure equations from \refe{eq:stressinpolar} with $\psi\sim r^2 F(\theta)$ gives
\begin{align}
 \pfrac{p}{r} = \frac{1}{r}(F'''+4F'), \qquad \pfrac{p}{\theta} = 0,
\end{align}
and thus (at leading order in $r$) $p=p(r)$. To have a pressure without a singularity requires $F'''+4F'=0$, and this third order ODE has the general solution
\begin{align}
 F = B_1+B_2\sin(2\theta)+B_3\cos(2\theta), \label{eq:eqF}
\end{align}
where $B_i$ are constant coefficients.
From \refe{eq:eqF} and \refe{eq:bhedeg}, we thus impose stream function form
\begin{align}
 \psi \sim r^2(B_1+B_2\sin(2\theta)+B_3\cos(2\theta))&+r^3(C_1\sin\theta
+C_2\cos\theta+C_3\sin(3\theta)+C_4\cos(3\theta)),
\end{align}
keeping two orders in $r$ so as to obtain the leading order term in the pressure behaviour. We find after applying the boundary conditions (including \refe{eq:bctsgradsig}) with Navier-slip
\begin{gather}
\psi \sim -\frac{1}{4}\frac{Ur^2}{\lambda}\left( 1-\tan\theta_c\sin(2\theta)-\cos(2\theta) \right)
\qquad\qquad\qquad\qquad\qquad\qquad\qquad\qquad\nonumber\\\qquad\qquad\qquad
-\frac{1}{16}\frac{Ur^3}{\lambda^2}\left( \sin\theta- \tan\theta_c\cos\theta - \frac{(2\cos(2\theta_c)-1)\sin(3\theta)}{2\cos(2\theta_c)+1} + \tan\theta_c\cos(3\theta)\right),
\nonumber\\
u \sim \frac{1}{2}\frac{Ur}{\lambda}
\left( \tan\theta_c\cos(2\theta) - \sin(2\theta) \right)
,
\qquad
v \sim
\frac{1}{2}\frac{Ur}{\lambda}\left( 1-\tan\theta_c\sin(2\theta)-\cos(2\theta) \right)
,\nonumber\\
T_{rr} \sim
\frac{U}{\lambda}\left( \tan\theta_c\cos(2\theta) - \sin(2\theta) \right)
,\qquad
T_{r\theta} \sim
-\frac{U}{\lambda}\left( \tan\theta_c\sin(2\theta)+\cos(2\theta) \right)
,\nonumber\\
T_{\theta\theta} \sim -T_{rr}
,\qquad
p \sim -\frac{1}{2}\frac{Ur}{\lambda^2}\frac{\cos(\theta_c-\theta)}{\cos\theta_c},
\end{gather}
displaying the mathematical possibility of such a flow. We note here as part of the result the form of the surface tension was forced to satisfy
\begin{align}
 \pfrac{\sigma}{r} \sim \frac{U}{\lambda} \quad \Rightarrow \quad \sigma \sim \sigma_c + \frac{Ur}{\lambda}, \qquad\mbox{as }r\to0,
\end{align}
which gives a reasonable result for the surface tension $\sigma$, that it is a constant to leading order when $r\ll1$, the regime in which we are interested. This also gives a motivation as to why a model which allows for variable surface tension, \ie surface tension relaxation, may be able to predict a flow which has no logarithmic pressure singularity, but finite and non-zero stresses at the contact line. Such surface tension relaxation is a feature of the interface formation model \cite{ShikhBook}, where the surface tension is given as $\sigma=\sigma(\rho^s)$ and the surface density is a function of position and time, \ie $\rho^s=\rho^s(\tbf{x},t)$.

\section{An intermediate region for quasistatic spreading}
\label{app:inter}

Motivated by the analysis of \cite{Hocking83} and the matching conditions \refe{eq:innmatch}, we introduce the intermediate scalings
\begin{align}
\Phi = C_a\xi\phi, \qquad \hat\zeta = \hat\epsilon\ln\left(e^{C_a^2C_b}C_a\xi\right), \label{eq:intscals}
\end{align}
where $\hat\epsilon = 1/|\ln\delta|$. Neglecting exponentially small terms, \refe{eq:ndlwin} transforms to
\begin{align}
\pfrac{}{\hat\zeta}\left\{ e^{\hat\zeta/\hat\epsilon-C_a^2C_b}\left[ C_a\dot{a}\phi
+ C_a^4\phi^3\left( \hat\epsilon^3\pppfrac{\phi}{\hat\zeta}-\hat\epsilon\pfrac{\phi}{\hat\zeta} \right) \right] \right\}=0,
\end{align}
so that to leading order in $\hat\epsilon$ we have
\begin{align}
 \phi = \left( \mathcal{K} + \frac{3\dot{a}\hat\zeta}{C_a^3\hat\epsilon} \right)^{1/3},
\label{eq:intmpower}
\end{align}
where $\mathcal{K}$ is a constant of integration. Expanding these for $\mathcal{K} \gg \ilfrac{3\dot{a}\hat\zeta}{(C_a^3\hat\epsilon)}$ and rewriting in inner variables, we have
\begin{align}
\label{eq:intmatch}
 \pfrac{\Psi}{\xi} \sim C_a\mathcal{K}^{1/3} + \dot{a}\mathcal{K}^{-2/3}\lrsq{C_b + \ln(\mye\, C_a \xi)/C_a^2},
\end{align}
which gives $\mathcal{K}=1$ to match with \refe{eq:innmatch}. To match this intermediate solution to the outer solution, \refe{eq:intmpower} clearly suggests matching should be performed with the cube of the free surface slope, indeed we find from
\refe{eq:intmpower} using scalings \refe{eq:intscals} (with our result that $\mathcal{K}=1$) that
\begin{align}
 -\lr{\pfrac{h}{x}}^3 \sim C_a^3 + 3\dot{a}\lr{C_a^2 C_b + \ln\lrsq{\frac{\mye\, C_a(a-x)}{\delta}}},
\end{align}
which is the same result as would be found directly from the inner region behaviour \refe{eq:innmatch}. From the outer region behaviour \refe{eq:outbehav} we have
\begin{align}
 -\lr{\pfrac{h}{x}}^3 \sim \frac{27}{a^6} + 3\dot{a}\ln\lrsq{\frac{\mye^3(a-x)}{2a}},
\end{align}
and thus the matching of these provides the spreading rate result given in \refe{eq:srr}.

\end{document}